\documentclass[aps,preprint,notitlepage]{revtex4-1}
\usepackage{graphicx}

\begin{document}

\title{Light enhancement by quasi-bound states in the continuum in dielectric arrays}
\author{Evgeny N. Bulgakov}
\author{Dmitrii N. Maksimov}
\affiliation{Kirensky Institute of Physics, 660036, Krasnoyarsk, Russia}

\begin{abstract}
The article reports on light enhancement by structural resonances
in linear periodic arrays of identical dielectric elements. As the
basic elements both spheres and rods with circular cross section
have been considered. In either case, it has been demonstrated
that high-$Q$ structural resonant modes originated from bound states
in the continuum enable near-field amplitude enhancement by factor
of $10$--$25$ in the red-to-near infrared range in lossy silicon. The asymptotic behavior of the $Q$-factor with the
number of elements in the array is explained theoretically
by analyzing quasi-bound states propagation bands.
\end{abstract}


\maketitle

\section{Introduction}

Light trapping by periodic structures is a mainstream idea in thin
film photovoltaics \cite{Yu12,Burresi15}. The idea dates back to the
paper by Sheng, Bloch, and Steplemann who fist proposed to employ
a periodic grating substrate in amorphous silicon solar cells
\cite{Sheng83}. Since then the periodic structures that diffract
light and increase the optical length within the absorbing
material have been a subject of numerous publications
\cite{Heine95,Bermel07,Kroll08,Wang13,Dhindsa16}. Most remarkably the
dielectric gratings have allowed for large enhancement factors
\cite{Yu10} which exceed the fundamental limit for a Lambertian
cell \cite{Yablonovitch82} that holds well only if there is a high density of
optical modes in the plane of the structure. At the same time as one approaches the nanophotonic
regime the wave nature of light comes into play to call for theoretical approaches taking into
account optical resonances \cite{Yu11}.

Due to numerous applications \cite{John12} trapping, focusing, and
concentration of light at the nanoscale have recently emerged as
topics of great interest primarily in regard to various plasmonic
devices \cite{Bartal09, Schuller10, Fang11, Pasquale12, Zhang15,
Jeon16, Zhang16} and photonic band-gap structures
\cite{Chutinan08, Lin09, Wang14, Othman16}. Another promising directions of research is all-dielectric
nanophotonics \cite{Savelev2015a, Jahani16} which employs
subwavelength dielectric objects, such as e.g. spherical
nanoparticles \cite{Tribelsky15}, to tailor the resonant response
of the system. Scattering by clusters dielectric spheres has been
in research focus for a long time
\cite{Borghese85,Ioannidou95,Xu95,Merchiers07,Wheeler2010a} with
high-$Q$ resonant modes in linear
\cite{Burin2004, Blaustein2007, Gozman2008} as well as
circular \cite{Burin06} arrays predicted a decade ago.
Nowadays, thanks to immense progress in manipulating dielectric
nanoparticles \cite{Fu13, Zywietz15,Dmitriev16}, we witness a surge of
interest in optical devices based on clusters and arrays of
dielectric spheres including subwavelength waveguides
\cite{Du2011,Savelev14, Bulgakov16}, optical nanoantenas
\cite{Krasnok2012, Fu13}, and circular oligomers \cite{Filonov14,
Chong14}.

In this article we consider light enhancement by high-$Q$ structural
resonances in linear one-dimensional periodic arrays of identical
dielectric elements. In what follows the term {\it structural resonance} will be applied to a resonance
whose position and width are dictated by the spatial distribution of dielectric elements. Thus, the structural
resonances can be contrasted to both natural material resonances and Mie resonances of individual dielectric elements.
As the basic elements both spheres and rods
with circular cross section will be considered. To achieve high-$Q$
resonances and, consequently, high enhancement factors we propose
to tune the parameters of the arrays to bound states in the
continuum (BSCs) which are localized states with the
eigenfrequency embedded into continuum of propagating solutions
\cite{Hsu16}. With the state of the art in the BSC research
thoroughly reviewed in the above reference here we simply
mention the key experiments on optical BSCs in periodic
dielectric structures \cite{Plotnik11, Lee12, Weimann13,
Chia_Wei_Hsu13, Corrielli13, Regensburger13}.

The BSCs in linear
periodic arrays of dielectric cylinders were predicted
theoretically in the earlier paper \cite{Bulgakov14} where the
role of the BSCs in wave scattering was also highlighted. More
recently, the problem of BSCs in arrays of dielectric rods
\cite{Yuan2016} was revisited with the existence domains in the
plane of radius and dielectric constant of the cylinders determined
through extensive numerical sumulations.
Unlike the guided solution below the line of light
\cite{Luan06,Du09} the qusi-bound modes in the parametric vicinity of a true BSC are resonantly coupled with
the freely propagating plane waves to emerge in the scattering cross-section as sharp Fano
resonances. In fact the collapse of a Fano resonance is a generic
feature inherent to BSCs \cite{Kim99,Venakides03,Guevara2003,Hein2012} including quasi-BSCs in
plasmonics \cite{Zou04}. In the
field of optics the above feature has allowed for design of narrow band
normal incidence filters \cite{Foley14,Foley15,Cui16}. The BSCs in linear arrays of dielectric spheres were first reported in \cite{Bulgakov15}.
As before for arrays of cylinders we contrast the
BSCs in arrays of spheres to the guided solutions below the line of light
\cite{Shore2004,Burin2004,Blaustein2007,Du2011,Linton13,Savelev14}. The effect of BSCs
on plane wave scattering was considered  \cite{Bulgakov16a, Bulgakov17} with the trapping of light with
orbital angular momentum theoretically predicted. Light enhancement by symmetry protected BSCs in photonic crystal slab
was demonstrated in \cite{Romano15, Mocella15, Yoon15}. Here, along the same line we propose to employ arrays of {\it finite} number of dielectric
elements. Although formally BSCs do not exist in finite structures \cite{Silveirinha14} their traces could be observed in form of high-$Q$ structural
resonances \cite{Bulgakov16} similarly to the traces of photonic band structure emerging in scattering on small clusters of spherical particles
\cite{Yamilov03}.

The article is organized as follows. In Section \ref{S2} we briefly review the BSCs in infinite arrays in dielectric rods.
Then we demonstrate structural resonances in the parametric vicinity of BSCs in finite arrays. A simple theory predicting the positions
of the structural resonances is proposed. In Section \ref{S3} we show
that the same theory applies for the arrays of dielectric spheres. In Section \ref{S4} we elaborate asymptotic behavior of the $Q$-factor and the enhancement
factor for various types of the structural resonances with the increasing number of elements in the array. Finally, we conclude in Section \ref{S5}.

\section{Quasi-BSCs in arrays of dielectric rods}\label{S2}

\begin{figure}
\includegraphics[width=1.\textwidth,trim={2.5cm 5cm 5cm
5cm},clip]{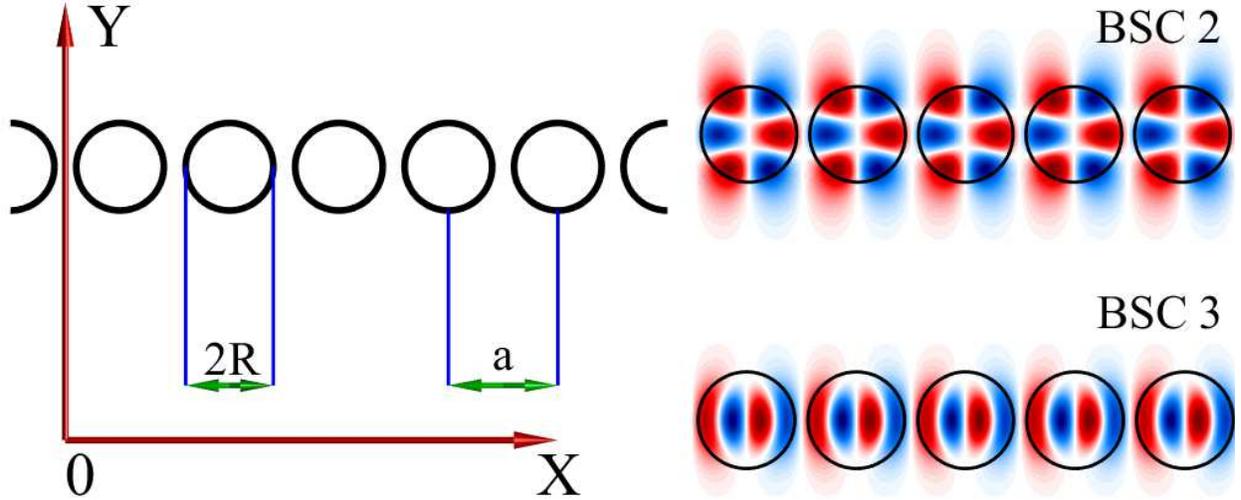} \caption{ BCSs in arrays of dielectric rods. Left panel: set-up of the array in
$x0y$-plane. Right panel: field patterns of $BSCs$ $2$, $3$ from Table \ref{Table1}.}\label{cylinders}
\end{figure}

We consider an array of identical dielectric rods of radius $R$ arranged along the $x$-axis
with period $a$. The axes of the rods are collinear and aligned with the $z$-axis. The cross-section of the array
in $x0y$-plane is shown in Fig. \ref{cylinders} (left panel). The numerical procedure for finding electro-magnetic
field in arrays of rods is well-known in literature \cite{Yasumoto05} and was adapted for BSCs in
\cite{Bulgakov14}. Suppose that the array is illuminated by a stationary TM plane wave with its electric vector pointing along
the axes of the rods, then according to the above references we have the scattered field inside the rods in the following form
\begin{equation}\label{cylinders_inside}
E_z(x,y)=\sum_{j}\sum_{m=0}^{\infty}c_{m}(j)e^{im\theta_j}J_m(k_0\sqrt{\epsilon}\rho_j),
\end{equation}
where $j$ enumerates the rods, $\theta_j, \rho_j$ are the local polar coordinates for the $j$ rod, $J_m(k_0\sqrt{\epsilon}\rho_j)$
- the Bessel function, $\epsilon$ - the dielectric permittivity, $c_{m}(j)$ - the expansion coefficient, and $k_0$ - the frequency.
Outside the rods we have a similar expression for the scattering function
\begin{equation}\label{cylinders_outside}
E_z(x,y)=\sum_{j}\sum_{m=0}^{\infty}\overline{c}_{m}(j)e^{im\theta_j}H_m^{(1)}(k_0\rho_j),
\end{equation}
where $H_m^{(1)}(k_0\rho_j)$ is the outgoing Hankel function. The summation over $m$ in Eqs. (\ref{cylinders_inside}) and (\ref{cylinders_outside})
runs to infinity. The only approximation in the numerical method is truncation to a finite number of summands in Eqs. (\ref{cylinders_inside})
and (\ref{cylinders_outside}). This approximation makes possible
to produce a finite interaction matrix of the scattering system and subsequently a finite number of equations for $c_{m}(j), \overline{c}_{m}(j)$
as described in \cite{Bulgakov14}. The method is known to converge rapidly with
the number of multipoles \cite{Yasumoto05}. In our computations we used $m=0,1,...,9$ which results in the relative truncation error $10^{-5}$ as defined
in \cite{Bulgakov16}. We refer the reader to \cite{Bulgakov14} for the lengthy equations for amplitudes $c_{m}(j), \overline{c}_{m}(j)$.

\subsection{BSCs in infinite arrays}
\begin{table}[t]
\begin{center}
\caption{BSCs in linear arrays of dielectric cylinders,
$\epsilon=15$. SP stands for symmetry protected. } \label{Table1}
\begin{tabular}{lcccccccc}
$BSC$ & SP & $k_0a$ & $k_x^0a$& $R/a$& $\nu$ & $\mu$ & $a_{\nu}$ & $a_{\mu}$ \\
\hline
1 & Yes & 1.8412 & 0& 0.4400 & 4 & 2 & 5.25 $\cdot10^{-4}$ & 0.20 \\
\hline
2 & Yes & 3.0758 & 0& 0.4400 & 2 & 2 & 9.6 $\cdot10^{-3}$ & 0.05 \\
\hline
3 & Yes & 3.5553 & 0& 0.4400 & 4 & 2 & 2.4 $\cdot10^{-3}$ & 0.177 \\
\hline
4 & Yes & 2.3864 & 0& 0.4400 & 2 & 2 & 6.85 $\cdot10^{-3}$ & 0.092 \\
\hline
5 & No & 2.8299 & 0& 0.4441 & 4 & 2 & 3.17 $\cdot10^{-3}$ & -0.085 \\
\hline
6 & No & 3.7156 & 1.6501 & 0.4400 & 2 & 1 & 5.27 $\cdot10^{-3}$ &  -0.0721 \\
\end{tabular}
\end{center}
\end{table}
In case of infinite arrays the summation over index $j$ in Eqs. (\ref{cylinders_inside}) and (\ref{cylinders_outside}) is run from minus to plus infinity.
The translation invariance allows to apply the Bloch theorem in the form \cite{Yasumoto05}
\begin{equation}\label{Bloch_cylinders}
c_m(j)=c_m(0)e^{ik_xaj}, \ \overline{c}_m(j)=\overline{c}_m(0)e^{ik_xaj},
\end{equation}
where $k_x$ is the $x$-axis component of the wave vector (Bloch vector).
Although the solution (\ref{cylinders_outside}) is
a superposition of outgoing functions it still can be totally decoupled from the far field outgoing waves due
to destructive interference between the waves emanating from different rods \cite{Bulgakov14}. In that situation a BSC
exists in the system even without the array being illuminated from the far zone. In dependence on
$k_x$ the BSCs are exceptional points in the quasi-BSCs propagation bands where the imaginary part of the resonant frequency
$k_0$ turns to zero \cite{Bulgakov16}. In general the asymptotic behavior of the imaginary and real parts of the resonant frequency in the vicinity of a BSCs is described by
the following formulas
\begin{equation}\label{imaginary}
-\Im\{k_0a\}=a_{\nu}(k_xa-k_x^{BSC}a)^{\nu}+\mathcal{O}[(k_xa-k_x^{BSC}a)^{\nu+2}],
\end{equation}
and
\begin{equation}\label{real}
\Re\{k_0a\}=k_0^{BSC}a-a_{\mu}{(k_xa-k_x^{BSC}a)}^{\mu}+\mathcal{O}[(k_xa-k_x^{BSC}a)^{\mu+1}],
\end{equation}
where $k_0^{BSC}$ is the BSC frequency, $k_x^{BSC}$ - the BSC wave
vector along the array axis, and $\nu, \mu$ - the leading
coefficient of polynomial expansion \cite{Bulgakov16}. The
physical meaning of Eq. (\ref{imaginary}) is simple; once the
system is detuned in $k_x$-space from a true BSC point $k_x^{BSC}$
the resonance acquires finite life-time given by the inverse of
$\Im\{k_0\}$. Although the majority of BSCs have $k_x^{BSC}=0$,
there are also so-called Bloch BSCs which are trapped waves
travelling along the array \cite{Bulgakov14, Bulgakov15, Gao16}.
In Table \ref{Table1} we collect the parameters of six BSCs found
for infinite arrays with $\epsilon=15$ (Silicon). The values of
$a_{\mu}$ and $a_{\nu}$ in Table \ref{Table1} were extracted from the data
by a polynomial fit. For a typical picture  of the band structure
of Bloch BSCs the reader is addressed to \cite{Bulgakov16}.

There are two generic types of BSCs in arrays of dielectric rods \cite{Bulgakov14}, symmetry protected, and unprotected by symmetry. Field patterns
of symmetry protected $BSCs\ 2, 3$ are shown in the left panel in Fig. \ref{cylinders} where one can see that those
BCSs are symmetrically mismatched with a plane wave normally incident to the array. The asymptotic behavior of the resonance live-times in
the vicinity of a BSC was recently considered by Yuan and Lu \cite{Yuan17} who argued that for symmetry protected BSCs the leading term should be
$\nu=2$. There were no, however, reliable estimate for $a_{\nu=2}$. In our numerical simulations we found that both $\nu=2$ and $\nu=4$ are possible.
In Table \ref{Table1} we presented the data on four symmetry protected BCSs a couple for each ${\nu=2}$ and ${\nu=4}$.
 The asymptotic behavior of unprotected BSCs was also considered in
\cite{Yuan17} with a mathematically rigorous result that for the the standing wave BSCs unprotected by symmetry the leading term is ${\nu=4}$ which
is in accordance with our findings for $BSC\ 5$. For Bloch $BSC\ 6$ we found ${\nu=2}$.

\subsection{Quasi-BSCs in finite arrays}
\begin{figure}[t]
\includegraphics[width=1.\textwidth,trim={0.8cm 0.0cm 1cm 0cm},clip]
{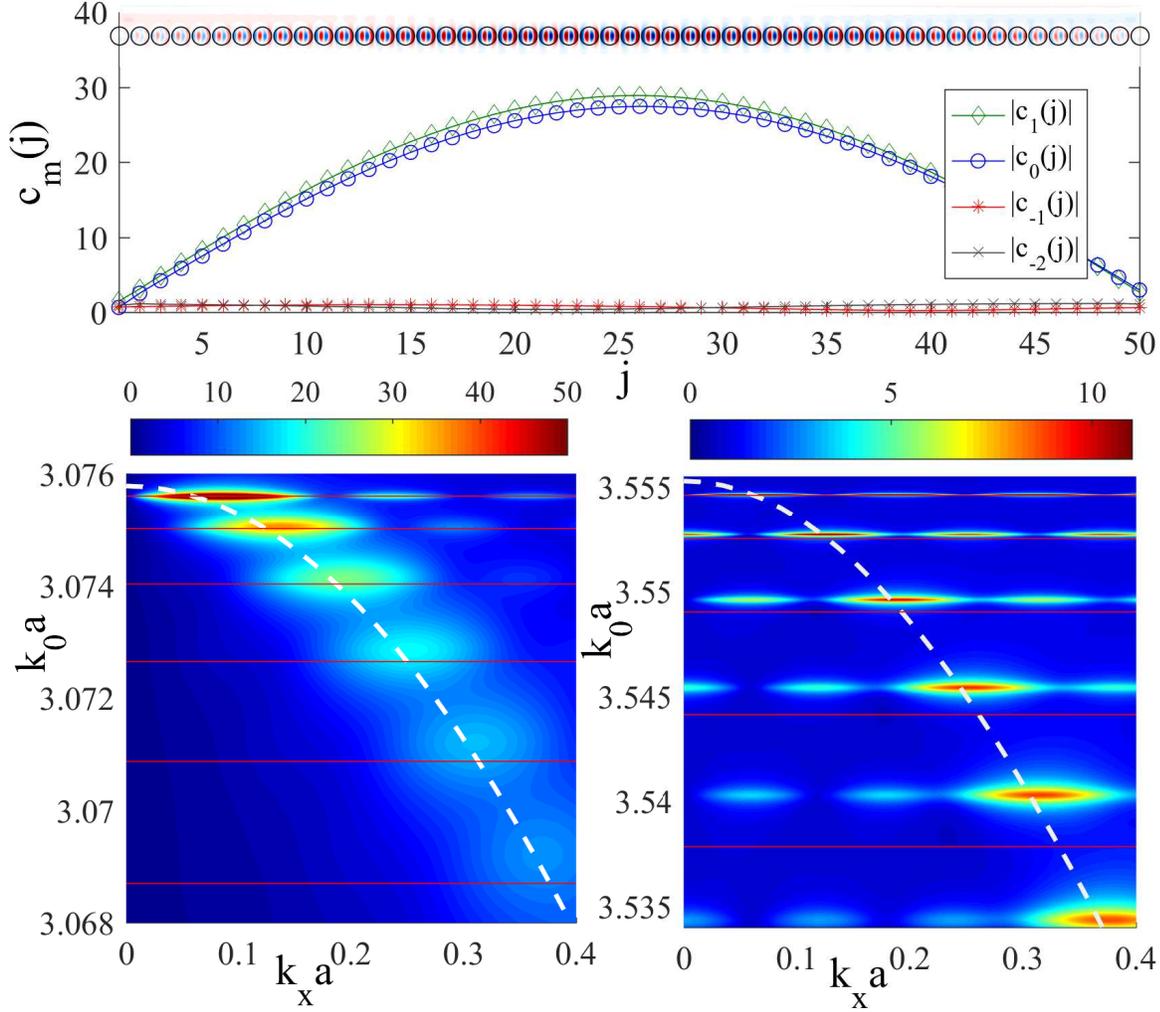} \caption{ Light scattering in the parametric vicinity of symmetry protected BSCs for dielectric $\epsilon=15$  arrays of rods
under illumination by a TM plane wave with unit amplitude.
Top panel: expansion coefficients $c_{m}(j)$ vs. the number of the rod $j$ for $N=50$ for $BSC\ 3$. The impinging wave parameters are tuned to the
first resonance Eq. (\ref{halfwave}).  The resulting field pattern
is shown on top of the subplot. South-west: mean value of the leading coefficient $\langle c_{m_{0}}\rangle$ vs. $k_0, k_x$ in the vicinity of $BSC \ 2$ for $N=50$.
South-east: The same for $BSC\ 3$. White dash lines correspond to asymptotic behavior by Eq. (\ref{real}). The frequencies of structural resonances by Eq. (\ref{spectrum}) are
shown by red horizontal lines.
}\label{Protected}
\end{figure}

Next we discuss the the light scattering by various types of quasi-BSCs in finite arrays
to highlight the features related to the asymptotic behavior described by Eqs. (\ref{imaginary}) and (\ref{real}). The array now consists of $N$ elements
and summation over $j$ in Eqs. (\ref{cylinders_inside}) and (\ref{cylinders_outside}) runs from $1$ to $N$.
In Fig. \ref{Protected} we present the results of numerical simulations of wave scattering by finite arrays of $N=50$ rods. The parameters $k_0, k_x$ which
are the frequency and the $x$-component of an impinging TM plane wave were swept
in the parametric vicinity of $BSCs\ 2$ and $3$. Notice that only positive $k_x$ were considered for the system is symmetric with respect to the inversion of
the $x$-axis. For simplicity the response was recorded as the mean of the absolute value of the leading coefficient in the expansion
 Eq. (\ref{cylinders_inside})
\begin{equation}\label{response}
\langle |c_{m_0}|\rangle =\frac{1}{N}\sum_{j=1}^N |c_{m_0}(j)|,
\end{equation}
where $m_0$ is the index of the coefficients $c_{m_0}(j)$ in Eq. (\ref{cylinders_inside}) with the largest absolute
value (see top panel in Fig. \ref{Protected} where $m_0=1$).

In both cases one can see "bright" spots following the asymptotic expression for the real part
of the resonant frequency (\ref{real}). These spots correspond to the structural resonances with the widths of the spots along $k_0$-axis
proportional to the inverse $Q$-factors which will be considered in Section \ref{S4}. The nature of the structural resonances
could be easily understood by close examination of the upper panel in Fig. \ref{Protected} where we plotted the expansion coefficients $c_m(j)$
for the first resonant spot. One can see from the upper panel in Fig. \ref{Protected} that the solution is nothing but a sinusoidal standing wave
locked between the edges of the array. Thus, to recover the spectrum of the structural resonances one can write down the following condition
for the $k_x^{(p)}$ of the $p$ structural resonance.
\begin{figure}[t]
\includegraphics[width=1.\textwidth,trim={1.2cm 0.0cm 1.5cm
2cm},clip]{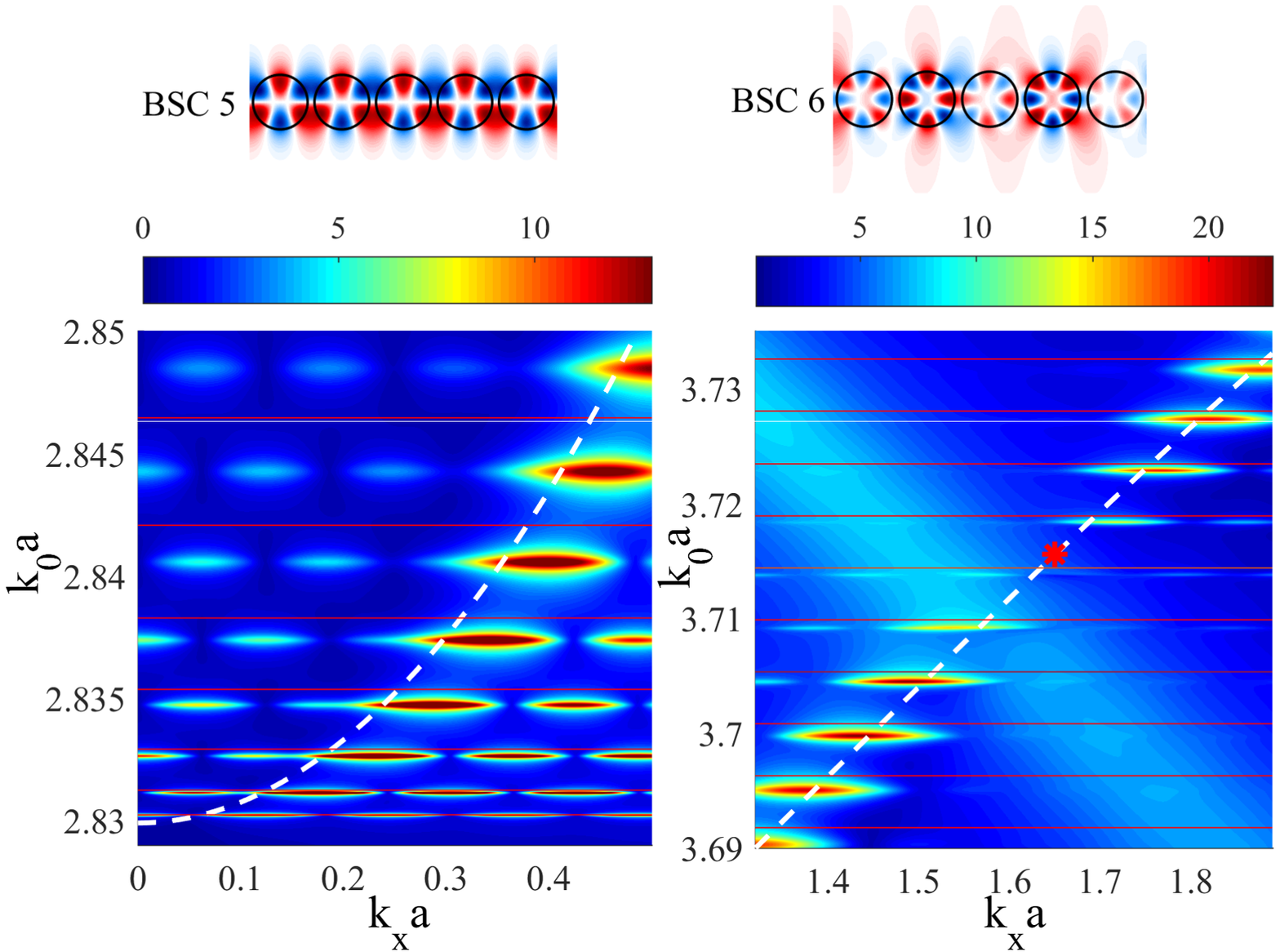} \caption{ Light scattering in the parametric vicinity of BSCs unprotected by symmetry for dielectric
$\epsilon=15$ arrays of rods under illumination by a TM plane wave with unit amplitude.
Left panel: Standing wave $BSC\ 5$ unprotected by symmetry; mean value of the leading coefficient $\langle c_{m_{0}}\rangle$ Eq. (\ref{response})
vs. $k_0, k_x$ in the parametric vicinity of the BSC for $N=50$ rods. The BSC field pattern is shown on top of the subplot.
Right panel: the same for Bloch $BSC \ 6$; the star shows the position of  $BSC\ 6$ in the parametric space $k_0, k_x$.
The BSC field pattern is shown on top of the subplot in form of the real part of the travelling wave amplitude.  White dash lines correspond to asymptotic
behavior by Eq. (\ref{real}). The frequencies of structural resonances by Eq. (\ref{spectrum}) are shown by red horizontal lines.}\label{Unprotected}
\end{figure}
\begin{equation}\label{halfwave}
Nak_x^{(p)}=\pi p, \ p=1,2,\ldots, .
\end{equation}
Then by virtue of Eq. (\ref{real}) we have
\begin{equation}\label{spectrum}
\Re\{k_0^{(p)}a\}=k_0^{BSC}a-a_{\mu}{\left(\frac{\pi p}{N}-k_x^{BSC}a \right)}^{\mu},
\end{equation}
with $p$ corresponding to the number of half-wavelengths between the edges and the field patterns
\begin{equation} \label{BSCpattern}
E_z^{(p)}(x)\simeq f(x,y)\sin\left(\frac{\pi px}{Na}\right),
\end{equation}
where $f(x,y)=f(x+a,y)$ is the periodic functions corresponding to the BSCs profiles in Fig. \ref{cylinders}. Remarkably, despite the
BCSs are standing waves the position of the first structural resonance $p=1$ is shifted from $k_x=0$ according to Eqs. (\ref{halfwave}) and
(\ref{spectrum}).
The field pattern of the
first resonance is shown on top of the upper panel in Fig. \ref{Protected} to visualize Eq. (\ref{BSCpattern}). Notice the
resemblance with $BSC\ 3$ in Fig. \ref{cylinders}. The resonant frequencies by Eq. (\ref{spectrum}) are plotted in Fig. \ref{Protected}
by red horizontal lines. One can see that Eq. (\ref{spectrum}) predicts to a good accuracy the positions of two first resonances $p=1,2$. For $p\geq 3$
the positions of resonances deviate from Eq. (\ref{spectrum}) due to the higher order terms in the polynomial expansion Eq. (\ref{real}). This effect
could be eliminated by further increase of the length of the array $Na$ with all resonances shifting to the BSC point according to Eq. (\ref{spectrum}).
In addition to the main sequence (\ref{spectrum}) in Fig. \ref{Protected} one can see satellite resonant peaks. We
speculate that the satellites are due to oscillating coupling of the structural resonances to the impinging wave with the same frequency but
a different $k_x$ component. Finally, one can see from Fig. \ref{Protected} that a BSC with $\nu=4$ produces a
picture of resonances with smaller high-intensity spots to evidence higher $Q$-factors.

In Fig. \ref{Unprotected} we present the results of numerical simulations of wave scattering by finite arrays of $N$ rods in the parametric vicinity of
$BSCs\ 5$ and $6$ from Table \ref{Table1}. The field patterns are shown on top of the plot to demonstrate that now the BCSs are not symmetrically
mismatched to the normally incident wave. In contrast to symmetry protected ones the BSCs of this sort are much more difficult to come by
for they always require some adjustment of either radius $R$ or Bloch vector $k_x$ to occur through an involved interference picture \cite{Bulgakov14}.
One can see from Fig. \ref{Unprotected} that the response function (\ref{response}) demonstrates features very similar to those in Fig. \ref{Protected}
with structural resonances emerging at frequencies by Eq. (\ref{spectrum}). Again for standing wave $BSC\ 5$ we see that only position of two first
structural resonances are accurately predicted by Eq. (\ref{spectrum}). Another source of error could be additional phase accumulated in
reflection from the edge of the array not accounted for by Eq. (\ref{halfwave}). Also notice that for Bloch $BSC\ 6$ the structural resonances are
equidistant with $a_{\mu=1}$ being the group velocity according to Eq. (\ref{real}).

\begin{figure}[t]
\includegraphics[width=1.\textwidth,trim={0cm 5cm 0cm
3.5cm},clip]{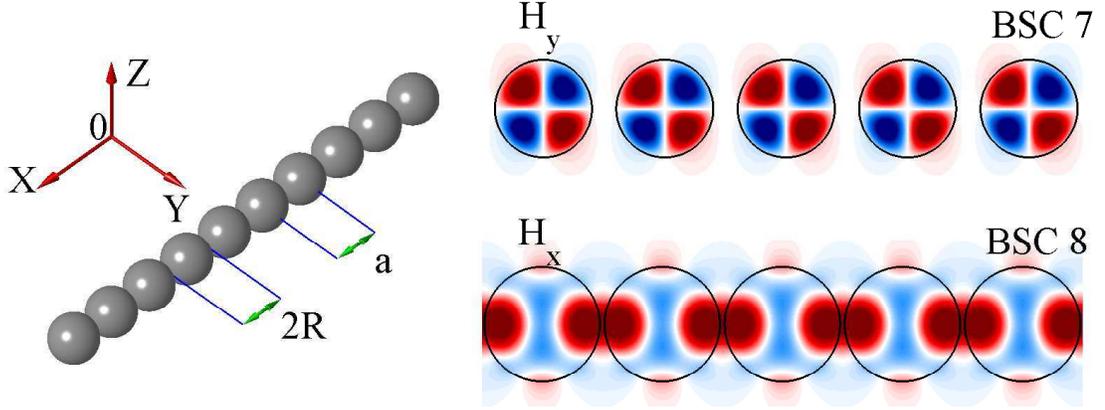} \caption{ BCS in arrays of dielectric spheres. Left panel: set-up of the array.
Right panel: The field patterns in form of $H_y$ component for $BSC\ 7$ and $H_x$ component for $BSC\ 8$ in $x0z$-plane.}\label{spheres}
\end{figure}

\section{Quasi-BSCs in arrays of dielectric spheres}\label{S3}
The system under scrutiny is schematically depicted in Fig. \ref{spheres}.
For arrays dielectric spheres we will follow the recipes from \cite{Bulgakov15} where the method by Linton, Zalipaev, and
Thompson \cite{Linton13}
was adapted for finding BSCs. The parameters of two BSCs, one symmetry protected, and one unprotected by symmetry are given
in Table \ref{Table2}. In the above referenced method the magnetic and electric vectors could be found in terms of Mie coefficients $a_n^m(j), b_n^m(j)$.
For instance, outside the spheres one has for the the scattered EM field electric vector ${\bf E}({\bf r})$
\begin{equation}\label{expansion}
{\bf E}({\bf r})=\sum_{j}\sum^{\infty}_{n=m^{*}}
\left[a_n^m(j){\bf M}^{m}_{n}({\bf r}-{\bf r}_j)+b_n^m(j){\bf N}^{m}_{n}({\bf r}-{\bf r}_j)\right],
\end{equation}
where $j$ the number of the sphere in the array, ${\bf r}_j$ - the coordinates of the $j$ sphere, m - azimuthal number, $m^{*}=max(1,m)$,
and ${\bf N}^{m}_{n}({\bf r}),{\bf M}^{m}_{n}({\bf r})$ are spherical
vector harmonics \cite{Stratton41}. Again in the case of infinite arrays we have according to the Bloch theorem
\begin{equation}\label{Bloch_spheres}
a_n^m(j)=a_n^m(0)e^{ik_xaj}, \ b_n^m(j)=b_n^m(0)e^{ik_xaj}.
\end{equation}
In this paper we restrict ourself we with $m=0$, though BSCs with higher orbital angular momentum are possible
\cite{Bulgakov16a, Bulgakov17}.
The field patterns
of BSCs from Table \ref{Table2} are shown in the left panel in Fig. \ref{spheres}.
Notice that $BSC$ 7 in Table \ref{Table2} is a wave of the TM-type which means $H_x$=0, and $b_n^{0}(j)=0$ \cite{Bulgakov15}. In contrast
$BSC$ 8 is a TE-wave with $E_x$=0, and $b_n^{0}(j)=0$.

\begin{table}[t]
\begin{center}
\caption{BSCs in linear arrays of dielectric spheres,
$\epsilon=15$. SP stands for symmetry protected. } \label{Table2}
\begin{tabular}{lcccccccc}
$BSC$  & SP & $k_0a$ & $k_x^0a$& $R/a$& $\nu$ & $\mu$ & $a_{\nu}$ & $a_{\mu}$ \\
\hline
7 & Yes & 3.6022 & 0 & 0.400 & 2 & 2 & 1.366 $\cdot10^{-2}$ & 0.010  \\
\hline
8 & No & 2.9280 & 0& 0.485 & 4 & 2 & 0.1818 $\cdot10^{-2}$ & -0.0793 \\
\end{tabular}
\end{center}
\end{table}

\begin{figure}
\includegraphics[width=1.\textwidth,trim={0cm 0cm 0cm
0cm},clip]{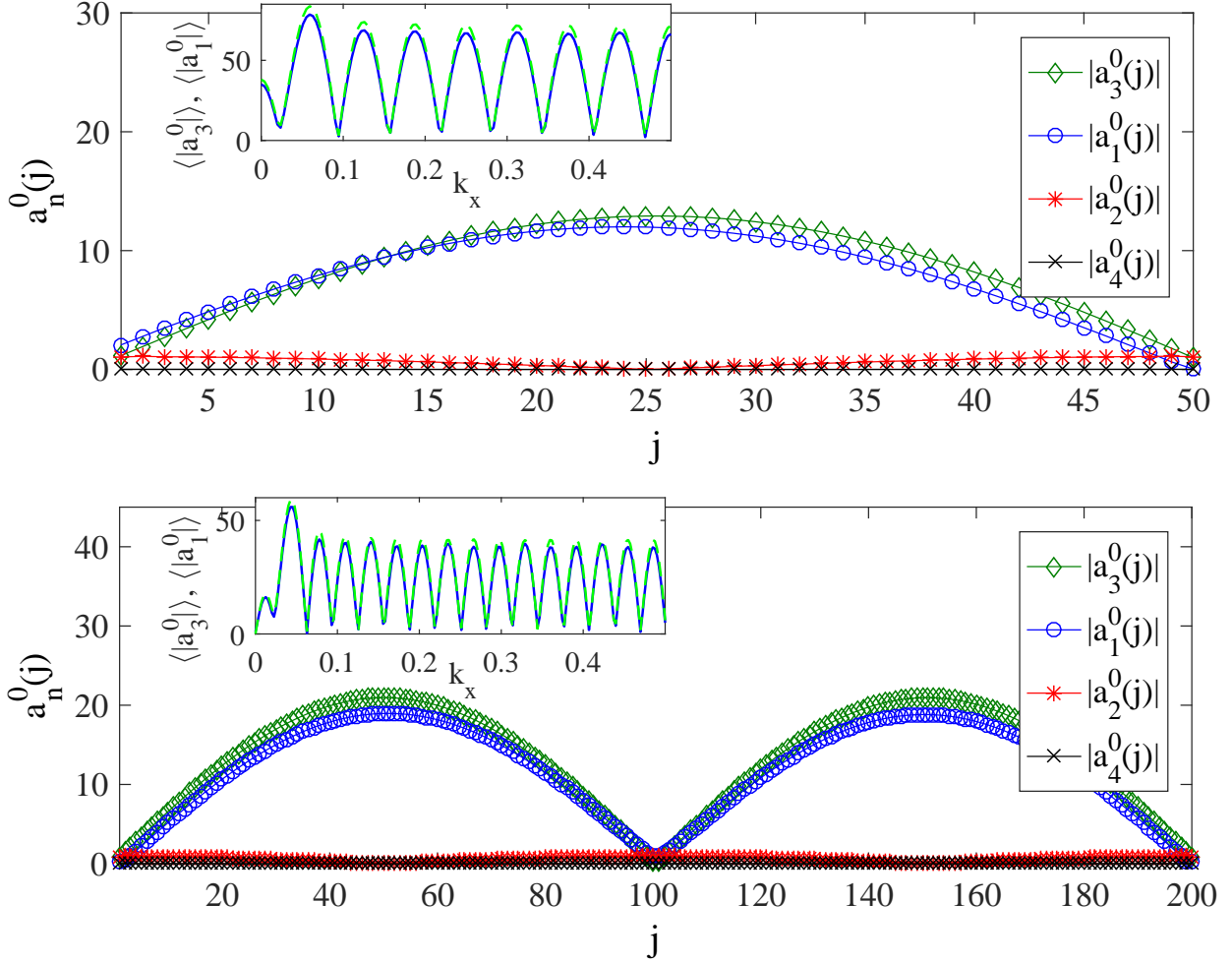} \caption{Light scattering by structural resonances in the parametric vicinity of $BSC\ 8$
in arrays of dielectric spheres. Top panel: the leading coefficients in Eq. (\ref{expansion}) for the first structural resonance $N=50$.
Bottom panel: the leading coefficients in Eq. (\ref{expansion}) for the second structural resonance $N=200$. The insets show the response function
Eq. (\ref{response}) evaluated with leading Mie coefficients $a^0_1(j), a^0_3(j)$ against the Bloch wave vector $k_x$. }\label{spheres_scattering}
\end{figure}

Let us now consider the resonant response of finite array with $N$ spheres.
Despite the similarity of Eq. (\ref{expansion}) to Eqs. (\ref{cylinders_inside}) and (\ref{cylinders_outside})
the numerical method is now computationally much more expensive due to the necessity to evaluate lattice sums with involved special functions \cite{Linton13}.
Though limited to the power of a desktop computer we can still circumvent the computational difficulties by recollecting the resonant picture
from the previous section. First by using Eq. (\ref{spectrum}) one can obtain the frequencies of a structural resonance. Then the response function
Eq. (\ref{response}) is computed numerically only in dependance on $k_x$ with coefficients $a_n^0(j), b_n^0(j)$ from Eq. (\ref{expansion}) now used
instead of $c_{m}(j)$ in Eq. (\ref{response}). The results for $BSC\ 8$ from Table \ref{Table2} under illumination by a TE-polarized plane wave of unit amplitude
are plotted in Fig. \ref{spheres_scattering}. Again one can see
 sinusoidal standing waves similar to Fig. \ref{Protected}. Notice that for the second resonance $p=2$ in Eq. (\ref{spectrum}) we intentionally chose
$N=200$ to eliminated the errors due to the higher order terms in Eq. (\ref{real}).
 The response function against $k_x$ for $k_0$ corresponding to the first and the
second resonances Eq. (\ref{spectrum}) is also demonstrated in the insets in Fig.
\ref{spheres_scattering}. Again one observes an oscillatory dependance with a pronounced maximum corresponding to Eq. (\ref{halfwave}) as in Figs.
\ref{Protected},\ref{Unprotected}. For $BSC\ 7$ the same results were observed under illumination by a TM-polarized plane wave.

\section{Light enhancement}\label{S4}
Bearing in mind the structure of resonant response of finite arrays we can now address the question of how many dielectric elements in the array
are needed to observe the effect of light enhancement in a realistic experiment with two figures of merit
being the quality and enhancement factors. Before proceeding to the assessment of those quantities we would like to mention that so far
all results were presented in dimensionless units to emphasize that the model is applicable to silicon dielectric arrays both in the visible
as well as in the near infrared where the real part of the dielectric constant varies insignificantly with the frequency of light \cite{Vuye1993,Green95}.
It should be noted, however, that the imaginary part in the same range can vary by orders of magnitude. Thus, one may expect that the effect of
light enhancement will be limited by the material losses in silicon.

\subsection{Q-factors}
\begin{figure}[t]
\includegraphics[width=1.\textwidth,trim={0cm 0cm 0cm
0cm},clip]{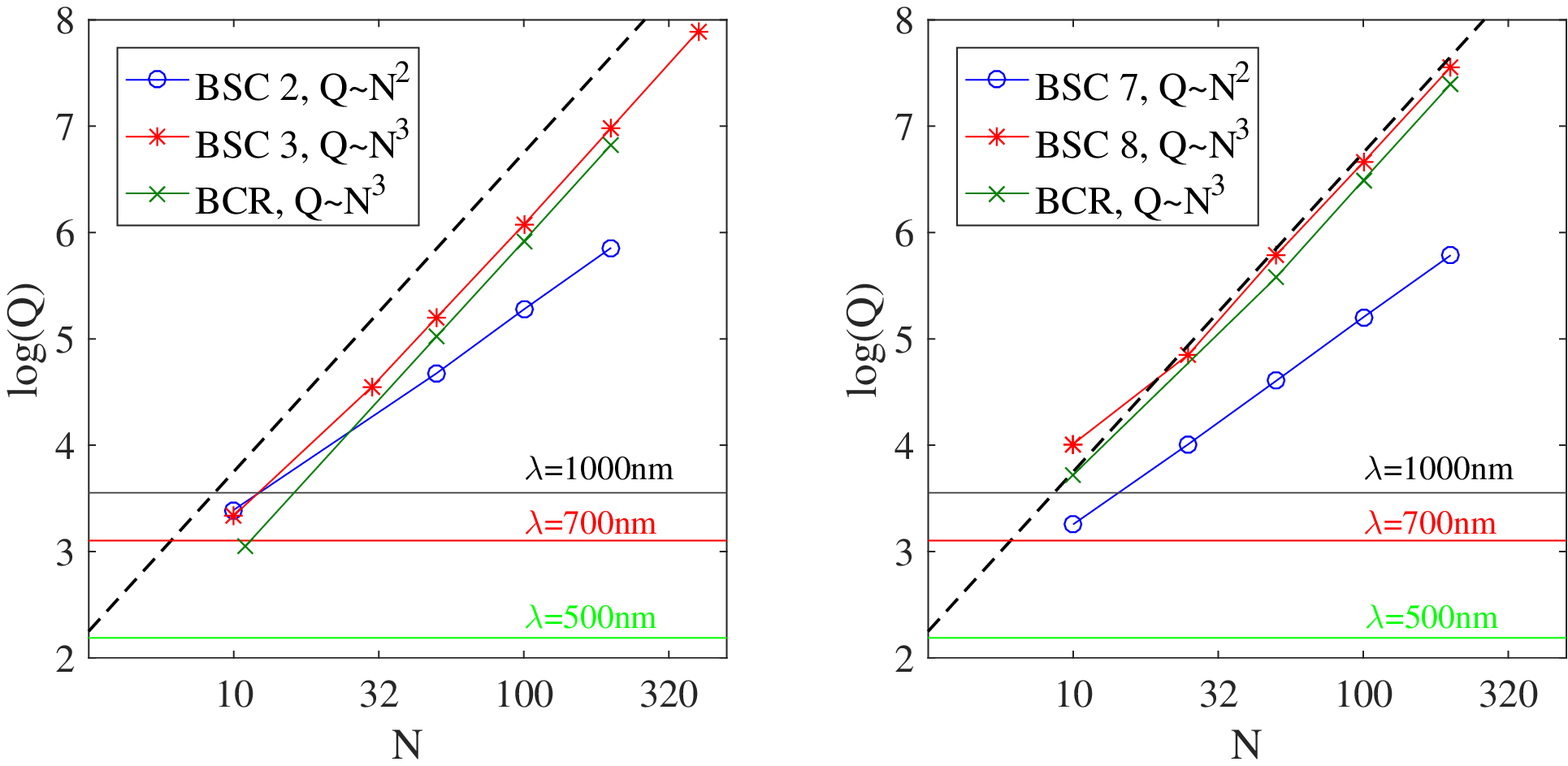} \caption{Q-factors for structural resonances against the number of dielectric elements $N$. Left panel: arrays of rods; BCR with $k_0a=2.8908,
k_xa=\pi, R/a=0.44$. Right panel: arrays of spheres; BCR with $k_0a=2.052,
k_xa=\pi, R/a=0.40$. The black dash line shows the best result found in \cite{Blaustein2007}. The horizontal
lines show the limits due to material losses in silicon.}\label{Q}
\end{figure}
Here we use the standard definition of the $Q$-factor
\begin{equation}
Q=\frac{\omega}{\gamma},
\end{equation}
where $\omega$ is the resonance frequency, and $\gamma$ is the
resonance width, both quantities extracted from the scattering data
described in the previous sections Figs. \ref{Protected},\ref{spheres_scattering}.
The $Q$-factors against the number of dielectric elements in the array
for the first ($p=1$ in Eq. (\ref{spectrum})) structural resonance of $BSCs\ 2,3,7$, and $8$ are plotted in Fig. \ref{Q}. As expected from Eqs.
(\ref{imaginary}) and (\ref{halfwave})
the $Q$-factor for resonances with $\nu=2$ scales as $Q\sim N^2$, however, for $\nu=4$ we observed $Q\sim N^3$ which, seemingly, contradicts our theoretical
predictions. The contradiction can be resolved by noticing that Eq. (\ref{imaginary}) only describes one specific mechanism of the radiation losses
equally applicable for both finite and infinite arrays, namely, the radiation from the sides of the arrays due to detuning from the true BSC point in
the parametric space. In the case of finite arrays there is another radiation mechanism due to the losses at the edges as the wave bounces between them
to form the sinusoidal resonant mode shape Eq. (\ref{BSCpattern}). Obviously, for the guided modes below the line of light that would be the only possible
radiation mechanism because the radiative losses at the sides are forbidden by total internal reflection. In fact, the scaling law for the edge radiation
is already obtained in \cite{Blaustein2007} where the authors demonstrated that the $Q$-factor scales as $Q\sim N^3$. Thus, one can conclude
that in case of $\nu=4$ the radiative losses at the sides are suppressed by the radiative losses at the edges resulting in the scaling law
for resonances formed by guided modes below the line of light \cite{Blaustein2007}.

In accordance with our findings we can now define {\it side-coupled resonances} with $Q\sim N^2$ for BSCs with $\nu=2$, and a wider class of
{\it edge-coupled resonances} with $Q\sim N^3$ which encompasses both BSCs with $\nu=4$ as well as the below continuum resonances (BCRs).
Now it is instructive to compare our results with the $Q$-factors for the BCRs. To do that in Fig. \ref{Q} we plot the $Q$-factors for two BCR
one for rods and spheres each. The parameters of the BCRs are given in the caption to Fig. \ref{Q}. One can see that both $\nu=2$ BSCs and BCR
demonstrate a similar dependance against $N$.

Finally, let us access the role of the material losses. The $Q$-factor limits were estimated based on the data on the imaginary part of refractive index
at $\lambda=500,700$nm by \cite{Vuye1993}, and at $\lambda=1000$nm by \cite{Green95}. One can easily see from Fig. \ref{Q} that $Q\approx 10^3$--$10^4$ resonances
can be observed with silicon in the red-to-near-infrared range for arrays of $N\approx 10$--$15$ elements.

\subsection{Enhancement factors}
\begin{figure}[t]
\includegraphics[width=1.\textwidth,trim={0cm 0cm 0cm
0cm},clip]{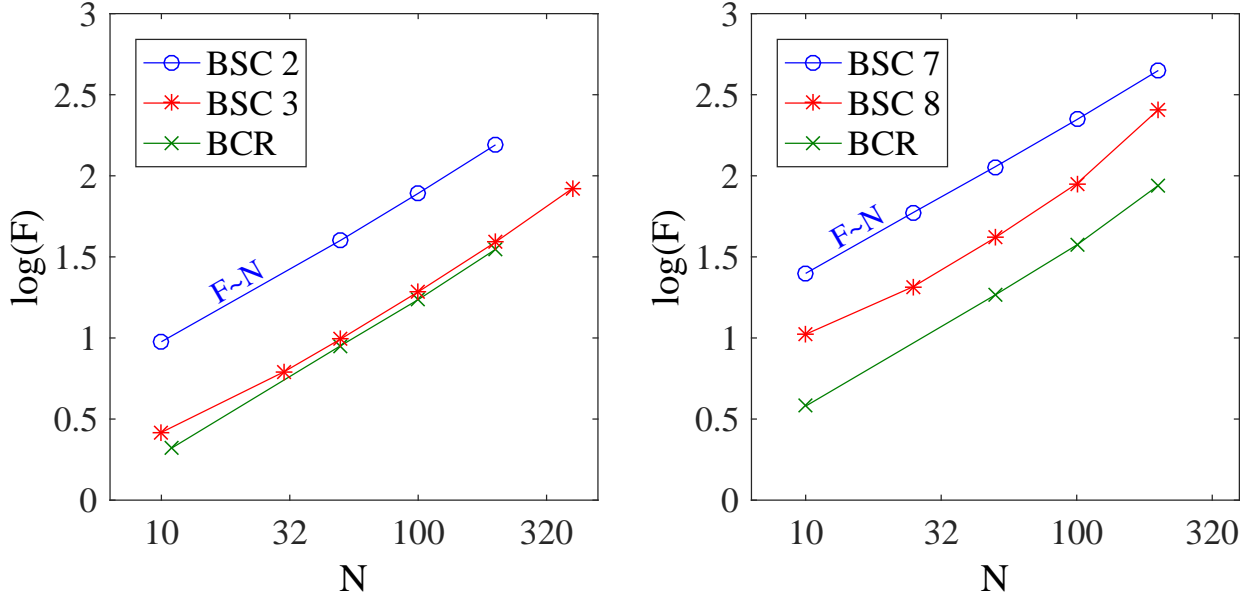} \caption{The enhancement factor of dielectric arrays. Left panel: arrays of rods. Right panel: arrays of spheres.
   }\label{F}
\end{figure}

We define the enhancement factor $F$ as the average of the square
root of the field intensity $I({\bf r})$ within the volume $V=\pi R^2Na$ (or area $2RNa$)
containing the whole of the finite array divided by the square root of the field
intensity carried by the incoming wave $I_0$.
\begin{equation}\label{Enhancement_Factor}
F=\frac{1}{\sqrt{I_0}V}\int_V d{\bf r} \sqrt{I({\bf r})}.
\end{equation}
Although according to Eq. (\ref{BSCpattern}) the EM-field varies
significantly in the vicinity of the array the enhancement factor
$F$ defined through Eq. (\ref{Enhancement_Factor}) is able to
quantify to what extent the field amplitude is enhanced by
resonant scattering. In Fig. \ref{F} we plot the enhancement factors
for the first ($p=1$ in Eq. (\ref{spectrum})) structural resonance of $BSCs\ 2,3,7$, and $8$
as well as the BCRs used in Fig. \ref{Q}. The arrays were illuminated with a plane wave with frequency equal to the
frequency of the resonance. The $k_x$ - component of the incident
wave each time was tuned to the maximum response which in case of BSC were given by Eq. (\ref{halfwave}). For BCR
the maximum response was found at the strict normal incidence.
One can see that only for the side-coupled resonances
$BSC$ 2,7 the enhancement factor is nicely fit by the expected dependence $F \sim \sqrt{Q} \sim N$
\cite{Yoon15}. Notice that in the above reference the enhancement was defined through intensity rather that amplitude.
For the edge-coupled resonances we were unable to produce a simple polynomial fit which is probably due to the residual side-coupling emerging
through the specific excitation of the resonance by a plane wave. More interesting, though,
is that the side-coupled resonances $BSC$ 2,7 produce significantly larger enhancement factors than the edge-coupled ones, and
hence should be opted for as the operating resonances for light enhancement.
In lossy silicon the limits for the enhancement factor of the side-coupled resonances will be placed at the same arrays lengths as
were found from Fig. \ref{Q} for the Q-factors. Thus, comparing Fig.\ref{F} against Fig.\ref{Q} one can conclude that the amplitude enhancement
factors $F\approx 10$--$25$ can be achieved with arrays of $10$--$15$ silicon subwavelength nanospheres in the red-to-near-infrared range with
the side-coupled resonances.

\section{Conclusion}\label{S5}

We demonstrated that the traces of bound states in the continuum in one-dimensional dielectric
arrays can be observed as a specific family of high-$Q$ resonances specified by the number of half-wavelengths
placed between the edges of the arrays. Two types of such resonances were
identified. The edge-coupled resonances possessing a higher $Q$-factor are similar to the below continuum resonances
known earlier \cite{Blaustein2007} to have the $Q$-factor scaling law as $Q\sim N^3$ with $N$ being the
number of elements in the array. In contract, the side-coupled resonances with lesser $Q$-factors have the scaling law $Q\sim N^2$.
Interestingly, it is the side-coupled resonances related to BSCs that under illumination by a plane wave demonstrate higher enhancement factors
providing a platform for light enhancement by finite dielectric arrays with the parameters tuned to a BSC.

It should be pointed out that in terms of both the $Q$-factor asymptotics and compactness our set-up is no match to
ultra-compact Fabry-Perot cavities \cite{Velha07} where the $Q$-factor scales exponentially with the number of periods in Bragg mirrors.
Those asymptotics, however, are hard to achieve in silicon nonophotonics as the subwavelength spheres are scaled down in size
to the visible wavelength where the material losses become significant. Our simulations show that in the red-to-near-infrared range
the $Q$-factors of $10^3$--$10^4$ and the amplitude enhancement factors $F\approx 10$--$25$ can be achieved with arrays of $10$--$15$ silicon subwavelength
nanospheres.

In contrast to familiar Fano resonances \cite{Kim99,Venakides03,Guevara2003,Hein2012} the structural resonances in finite systems are shown to
generate a complicated picture in the resonant response to illumination by a plane wave. With the frequency
of the impinging wave tuned to the structural resonance we observed the oscillating behavior of the enhanced EM-field amplitude under
variation of the wave vector component directed along the array. In real systems the effect of finiteness will be always eventually
obscured by material losses. One may expect that with the increase of the length of the array the many pronounced resonances will merge into a Fano feature.
We speculate that this picture could be explained by a properly set coupled mode theory \cite{Suh04} based on analytical estimates for
resonant frequencies and mode shapes obtained in this paper. Along with asymptotic scaling law for edge-coupled resonances enhancement factor this
constitutes a goal for future studies.

The bound states in the continuum were recently employed for engineering high-$Q$ resonators for compact nanophotonic lasers
\cite{Kodigala17} providing access to new coherent sources with intriguing topological properties for optical trapping, biological imaging,
and quantum communication.
To provide useful guidelines for practical implementations of structures supporting the bound states the effects of both structural
fluctuations \cite{Ni17}
and substrate coupling  \cite{Sadrieva17} have recently been considered.
As it was mentioned earlier in the introduction the BSCs are only allowed in infinite systems \cite{Silveirinha14}
so that in design of BSC-based devices the researchers will always be bound with residual radiation losses. We believe that our model of the BSC related
structural resonances sheds some light onto important aspects of emerging BSC nanophotonics \cite{Hsu16, Rybin17}.

\section*{Funding} This work was supported by Ministry of Education and Science of Russian Federation
(state contract N 3.1845.2017) and RFBR grant 16-02-00314.

\section*{Acknowledgment} We appreciate discussions with Almas F. Sadreev and Polina N. Semina.

\bibliography{Subwavelength_waveguides}

\end{document}